\documentclass{article}

\PassOptionsToPackage{authoryear,round}{natbib}

 \usepackage[preprint]{neurips_2025}

\usepackage[utf8]{inputenc} 
\usepackage[T1]{fontenc}    
\usepackage{hyperref}       
\usepackage{url}            
\usepackage{booktabs}       
\usepackage{amsfonts}       
\usepackage{nicefrac}       
\usepackage{microtype}      
\usepackage{xcolor}         
\usepackage{graphicx}
\usepackage{wrapfig}
\usepackage{booktabs, tabularx, xparse}
\usepackage{multirow}

\usepackage{xcolor}
\definecolor{lightgray}{RGB}{230,230,230}

\usepackage{enumitem}
\usepackage{pdfpages}

\usepackage{listings}
\usepackage{xcolor}
\usepackage{fancyhdr}
\usepackage{tcolorbox}
\usepackage{enumitem}

\lstdefinestyle{pythonstyle}{
    backgroundcolor=\color{backcolour},   
    commentstyle=\color{codegreen},
    keywordstyle=\color{magenta},
    numberstyle=\small\color{codegray},
    stringstyle=\color{codepurple},
    basicstyle=\rmfamily\small,
    breakatwhitespace=false,         
    breaklines=true,                 
    captionpos=b,                    
    keepspaces=true,                 
    numbers=left,                    
    numbersep=3pt,                  
    showspaces=false,                
    showstringspaces=false,
    showtabs=false,                  
    tabsize=1
}

\title{From Idea to Co-Creation: A Planner--Actor--Critic Framework for Agent Augmented 3D Modeling}

\author{%
  Jin Gao\\
  Massachusetts Institute of Technology\\
  Cambridge, MA 02139 \\
  \texttt{gaojin@mit.edu}
  \And
  Saichandu Juluri\\
  Northeastern University\\
  Boston, MA 02115 \\
  \texttt{juluri.s@northeastern.edu}
}

\usepackage{graphicx,tabularx,multirow,booktabs,xcolor,array}
\definecolor{lightgray}{RGB}{230,230,230}

\begin{document}

\maketitle

\begin{abstract}
We present a framework that extends the Actor-Critic architecture to creative 3D modeling through multi-agent self-reflection and human-in-the-loop supervision. While existing approaches rely on single-prompt agents that directly execute modeling commands via tools like Blender MCP, our approach introduces a Planner–Actor–Critic architecture. In this design, the Planner coordinates modeling steps, the Actor executes them, and the Critic provides iterative feedback, while human users act as supervisors and advisors throughout the process. Through systematic comparison between single-prompt modeling and our reflective multi-agent approach, we demonstrate improvements in geometric accuracy, aesthetic quality, and task completion rates across diverse 3D modeling scenarios.
Our evaluation reveals that critic-guided reflection, combined with human supervisory input, reduces modeling errors and increases complexity and quality of the result compared to direct single-prompt execution. This work establishes that structured agent self-reflection, when augmented by human oversight and advisory guidance, produces higher-quality 3D models while maintaining efficient workflow integration through real-time Blender synchronization.
\end{abstract}

\section{Introduction}

Language-based AI agents have rapidly evolved from simple, prompt-driven tools into versatile systems capable of autonomous task execution, including reading, creating, and modifying files; debugging code; performing internet and database searches; and operating desktop applications. Recent interoperability standards such as the Model Context Protocol (MCP) and the Agent-to-Agent (A2A) protocol \citep{Ehtesham2025} further enhance these capabilities by standardizing tool usage, structuring information exchange, and enabling collaboration across platforms. Building on these extended agent capabilities, multi-agent systems assign specialized roles to agents that work together via structured communication, which shows enhanced scalability, accuracy, and resilience, especially in complex tasks\citep{han2025llmmultiagentsystemschallenges}.

While current large-scale neural networks can generate high-quality 2D and 3D frames or models via text-to-3D and image-to-3D prompts, single-shot generations often fall short of user expectations\citep{jiang2024surveytextto3dcontentsgeneration}. These outputs typically require manual human evaluation and iterative optimization to progressively approach the desired result\citep{xiang2025sel3dcraftinteractivevisualprompts}. In traditional 3D modeling workflows, humans decompose a model into components and organize them using non-destructive structures like groups, blocks and layers to maximize edit potential. However, mainstream text-to-3D models produce unified meshes, which are difficult to edit. An alternative method control traditional 3D modeling software using APIs, enabling full modeling flexibility including lighting, textures, and third party assets. Meanwhile, evaluating 3D models remains a challenge. Recent advances in multimodal vision-language models (VLMs) have enabled AI systems to interpret and evaluate visual inputs, offering new potential for automated model assessment \citep{maiti2025gen3devalusingvllmsautomatic}.

Inspired by the actor-critic architecture from reinforcement learning \citep{Konda2000, SuttonBarto2020}, we hypothesize to enhance the quality of creative 3D modeling by integrating planning, action, and critique agents within a multi-agent workflow. Building on this, we further investigate the potential of incorporating human feedback during the agents' workflow, to assess whether human-in-the-loop supervision can provide additional improvements in the modeling outcome.

\section{Related Work}

Recent work has explored using large language models as agents to directly create or manipulate 3D content. \textit{SceneCraft} \citep{hu2024scenecraftllmagentsynthesizing} introduced an LLM agent that plans a scene graph and then generates Blender Python scripts to construct complex scenes, letting a vision–language model self-critique the result. \textit{WorldCraft} \citep{liu2025worldcraftphotorealistic3dworld} pioneered a collaborative agent system for scene design. It uses a “coordinator agent” to manage two specialist LLM agents – \textit{ForgeIt}, focuses on constructing and customizing individual 3D objects, and \textit{ArrangeIt}, handles spatial layout. This approach outperforms prior single-agent methods in both versatility and user control.

Endeavors also focus on fine-tuning domain expert agents to improve control quality. \textit{BlenderLLM} \citep{du2024blenderllmtraininglargelanguage} is a domain-tuned LLM that converts natural language instructions into Blender’s python (bpy) scripts, improving the reliability of generated 3D scripts over base models. 

The general paradigm of \textit{LLM-as-Critic} has proven effective in improving generation quality. In language tasks, \textit{Self-Refine} \citep{madaan2023selfrefineiterativerefinementselffeedback} and \textit{DeepCritic} \citep{yang2025deepcriticdeliberatecritiquelarge} showed that iteratively critiquing and revising an LLM's own outputs yields higher performance gains over one-shot answers. In graphics generation, \textit{CoherenDream} \citep{jiang2025coherendreamboostingholistictext} uses a fine-tuned vision-language model (\textit{3DLLaVA-Critic}) to evaluate the alignment between a generated 3D scene and the textual prompt, achieving state-of-the-art text alignment in multi-object 3D generation. These works demonstrate that a critic agent can lead to higher-quality outcomes. 

Finally, integrating human feedback into the LLM loop addresses the “humanity” side of agentic systems. The instability of autonomous systems can struggle with subjective criteria like aesthetics or may fall into infinite loops or premature convergence, making human supervision necessary. \textit{Sel3DCraft} \citep{xiang2025sel3dcraftinteractivevisualprompts} replaces blind trial-and-error prompting with a guided visual interface, helping users identify better results from multi-batch generations.

\section{Methodology}

\subsection{System Overview}

We selected Blender as the backend platform for testing 3D modeling due to its open-source nature, comprehensive functionality, and robust Python API. Blender offers extensive support for model import/export, modeling operations, material editing, and animation workflows. Additionally, Blender-MCP, a rapidly evolving open-source project, enables scene inspection, automated screenshots, and execution of Python code directly within Blender. It also supports integration with several popular 3D model generation APIs and asset libraries (e.g., PolyHaven, Hyper3D and Rodin), making it suitable for implementing basic agent-environment interaction functionalities.

To enhance the interactivity between Blender and the agent framework, we chose to implement a web-based interface for both agent interaction and real-time 3D model visualization and editing. In this setup, Blender serves as the backend, responsible for core modeling functionalities.

We designed and implemented a real-time, bidirectional communication pipeline between Blender and a React-based frontend using react-three-fiber. The key components include:

(1) an extended MCP server and Blender plugin based on the Blender-MCP project,

(2) a WebSocket-based communication module enabling bidirectional exchange of geometry, lighting, and texture data between Blender and the frontend,

(3) a multi-agent engine powered by CopilotKit, with LangGraph serves as agent chain-of-thought framework, with OpenAI ChatGPT 4.1 as the base model,

(4) a unified user interface built with Next.js and React--Three--Fiber.

Together, these components form a full-stack, agent-driven interactive modeling system that enables seamless collaboration between AI agents, human users, and the 3D modeling backend.

\begin{figure}[hbt!]
    \centering
    \includegraphics[width=1\linewidth]{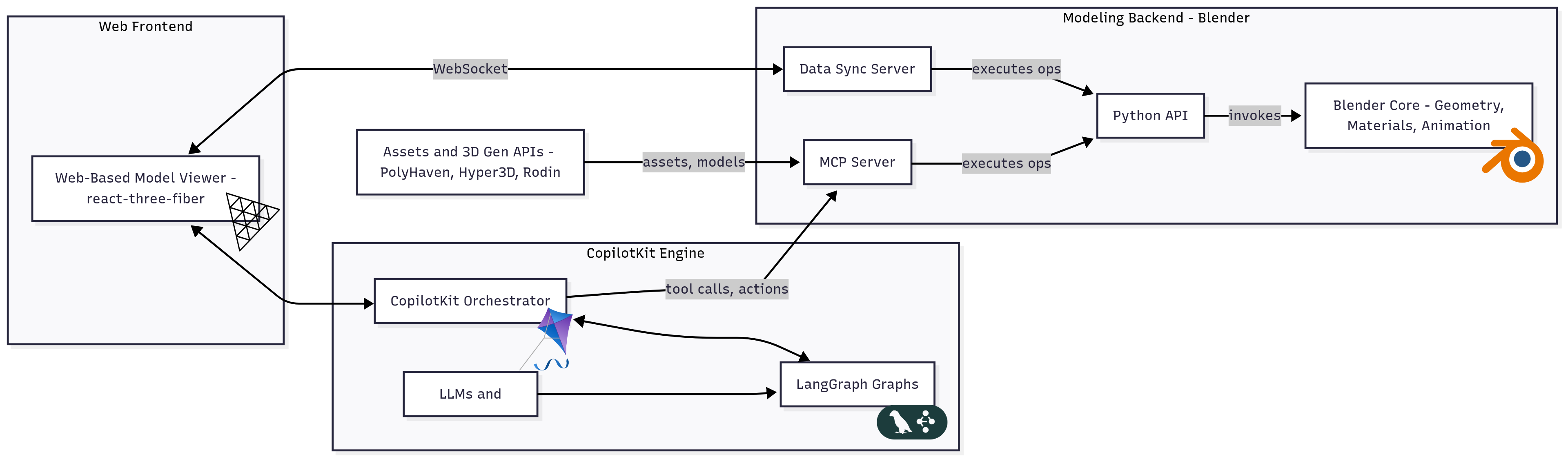}
    \caption{System Diagram}
    \label{fig:placeholder}
\end{figure}


\subsection{Agent Definition}

We decompose the modeling task into a collaborative workflow involving three specialized agents. This design is inspired by the workflow of coding agents, particularly Claude Code \citep{claudecode}, which maintains a structured plan within the context, maintain task continuity and mitigate issues such as context contamination and attention diffusion, which are common in long, multi-step generation tasks\citep{Farquhar2024}. In our multi-agent framework:

\begin{figure}[hbt!]
    \centering
    \includegraphics[width=0.6\linewidth]{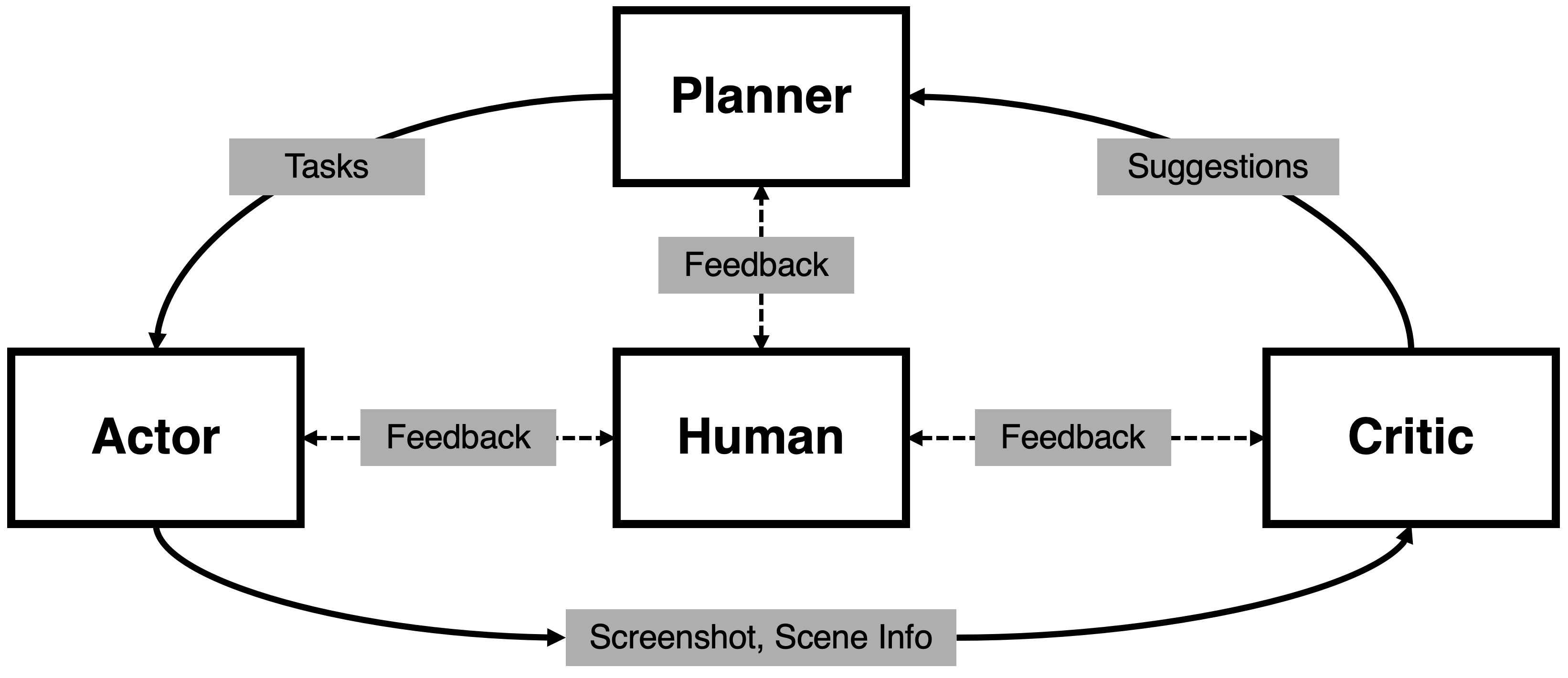}
    \caption{Agent-Human Co-creation Loop}
    \label{fig:placeholder}
\end{figure}

\textbf{Planner Agent} is responsible for proposing and maintaining a list of modeling steps (i.e., a todo list) based on the given task description. This list is persistently stored in a shared context and is continuously updated—steps can be added, modified, or removed based on the Actor's execution results and the Critic’s feedback.

\textbf{Actor Agent} autonomously selects a task from the todo list and executes the corresponding action using tools from a predefined toolset. These tools include operations such as primitive creation, modifier application, and model import from local or online libraries. While Blender-MCP already supports several MCP tools, we further extended its toolset to cover common modeling operations, such as creating basic geometry.

\textbf{Critic Agent} evaluates the current modeling result by analyzing the scene state and visual feedback (e.g., screenshots). It then generates structured critique and suggestions, which are fed back to the Planner Agent to edit the todo list.

In our prototype, the agent operates in a unidirectional sequence, progressing from planner to actor to critic, supported by a partially shared information board. A human participant may engage at any stage of the agent dialogue through a dedicated input interface, allowing them to direct, accept, or intervene in the generating process at the agent's action stage.

\subsection{Tooling and Integration}

We use Blender-MCP\citep{blender-mcp} as the backend system for real-time command execution in Blender. The current version of Blender-MCP supports a variety of MCP tools, including executing Python code, retrieving scene information, and downloading models from online 3D asset libraries. 

In our interface prototype, tools are organized into toolsets — such as modeling tools and refinement tools — that can be dynamically enabled, disabled, or switched during the modeling process. These toolsets are exposed to both agents and human users, allowing fine-grained control over the modeling workflow and enabling more precise human-agent collaboration.

\begin{figure}[hbt!]
    \centering
    
    \includegraphics[width=\linewidth]{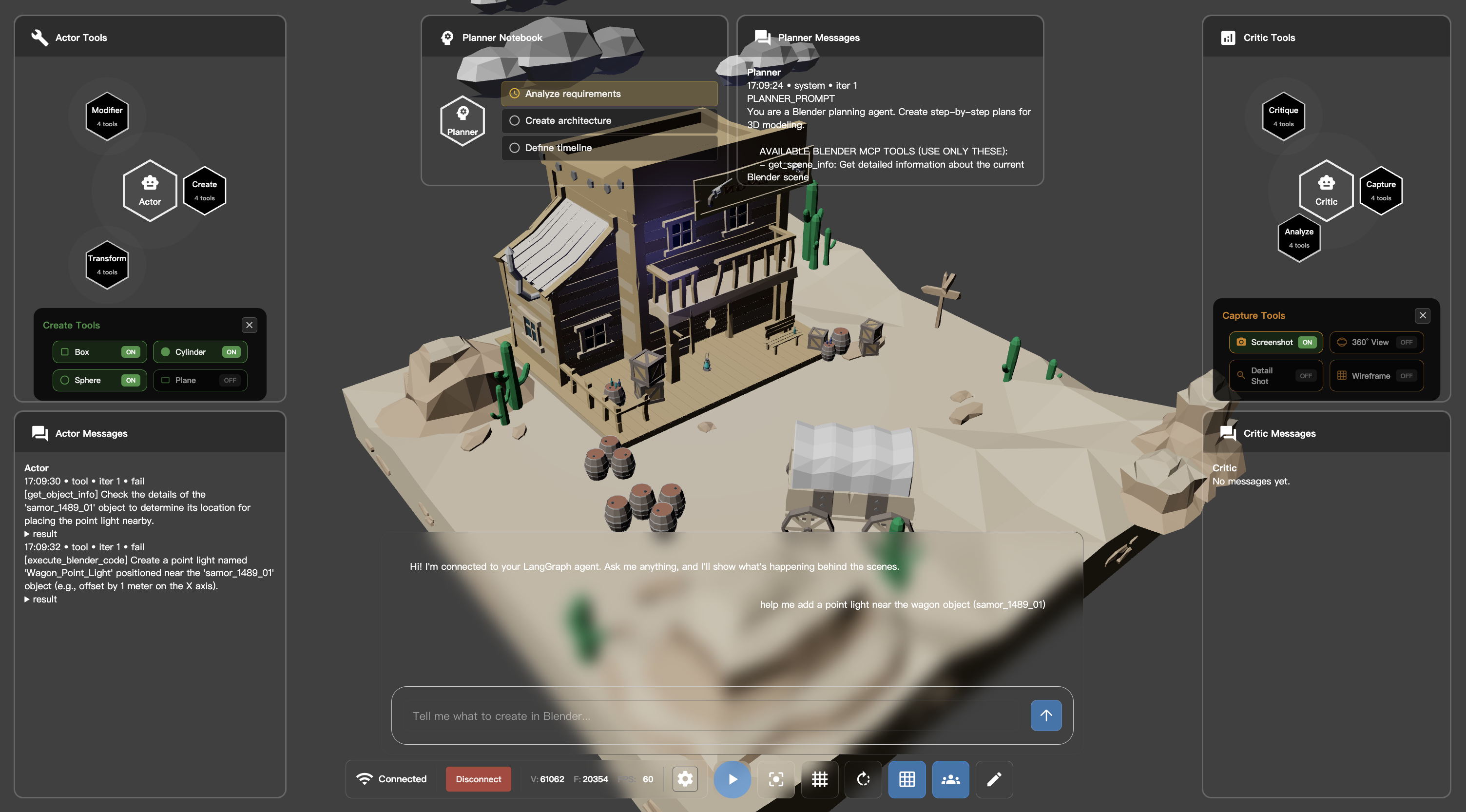}
    \vspace{0.7em} 
    
    \begin{minipage}[b]{0.32\linewidth}
        \centering
        \includegraphics[width=\linewidth]{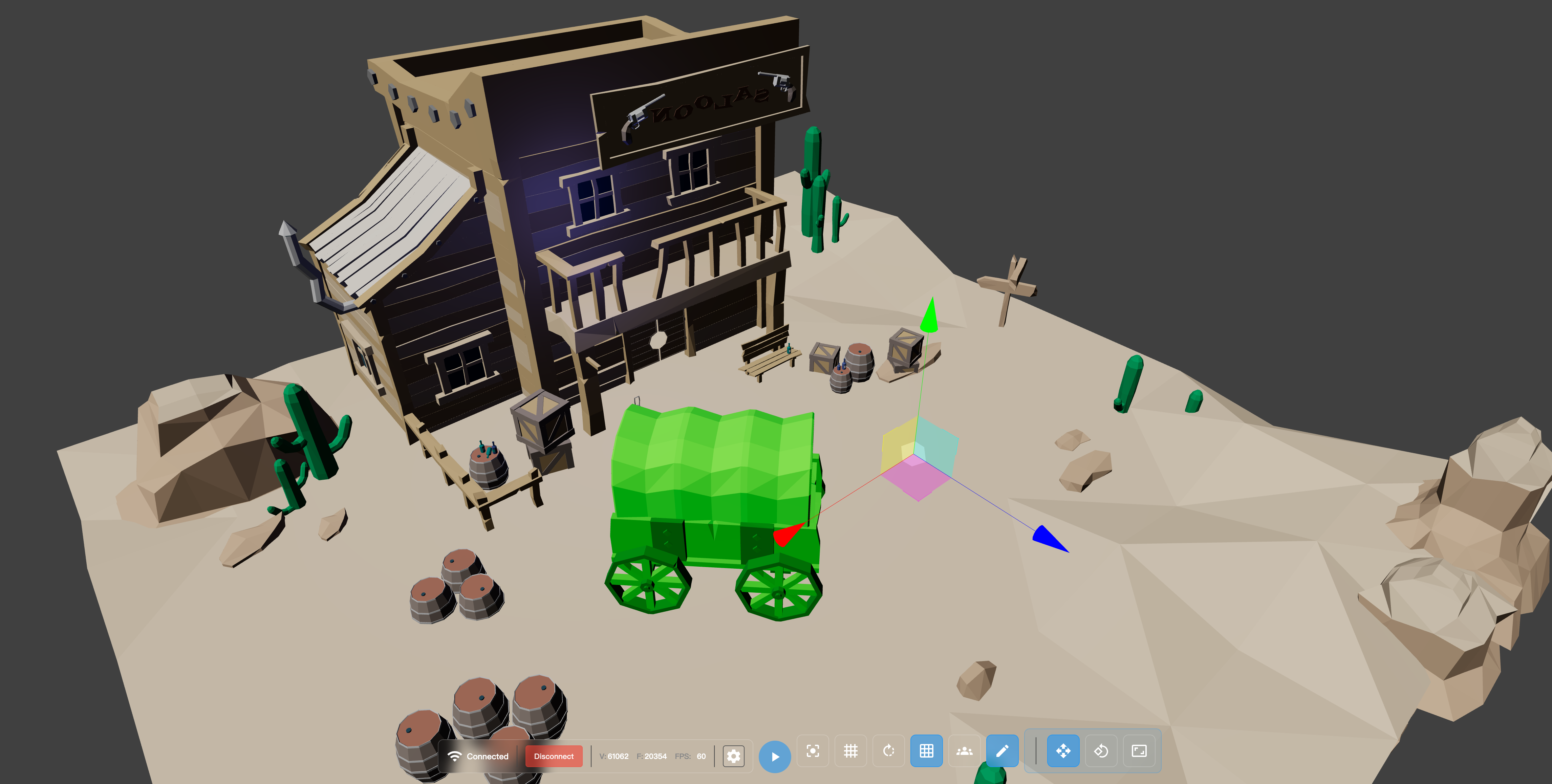}
    \end{minipage}
    \hfill
    \begin{minipage}[b]{0.32\linewidth}
        \centering
        \includegraphics[width=\linewidth]{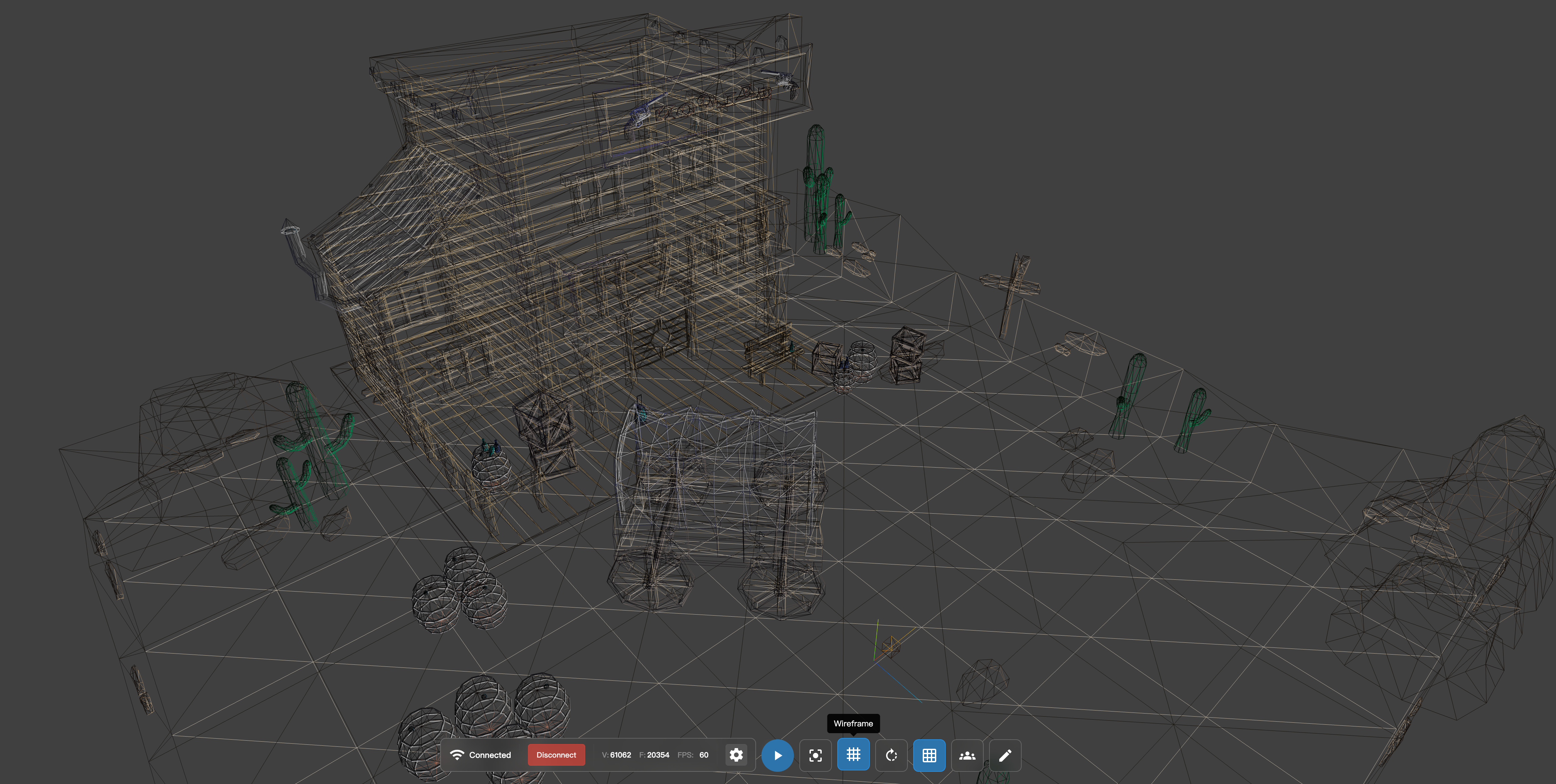}
    \end{minipage}
    \hfill
    \begin{minipage}[b]{0.32\linewidth}
        \centering
        \includegraphics[width=\linewidth]{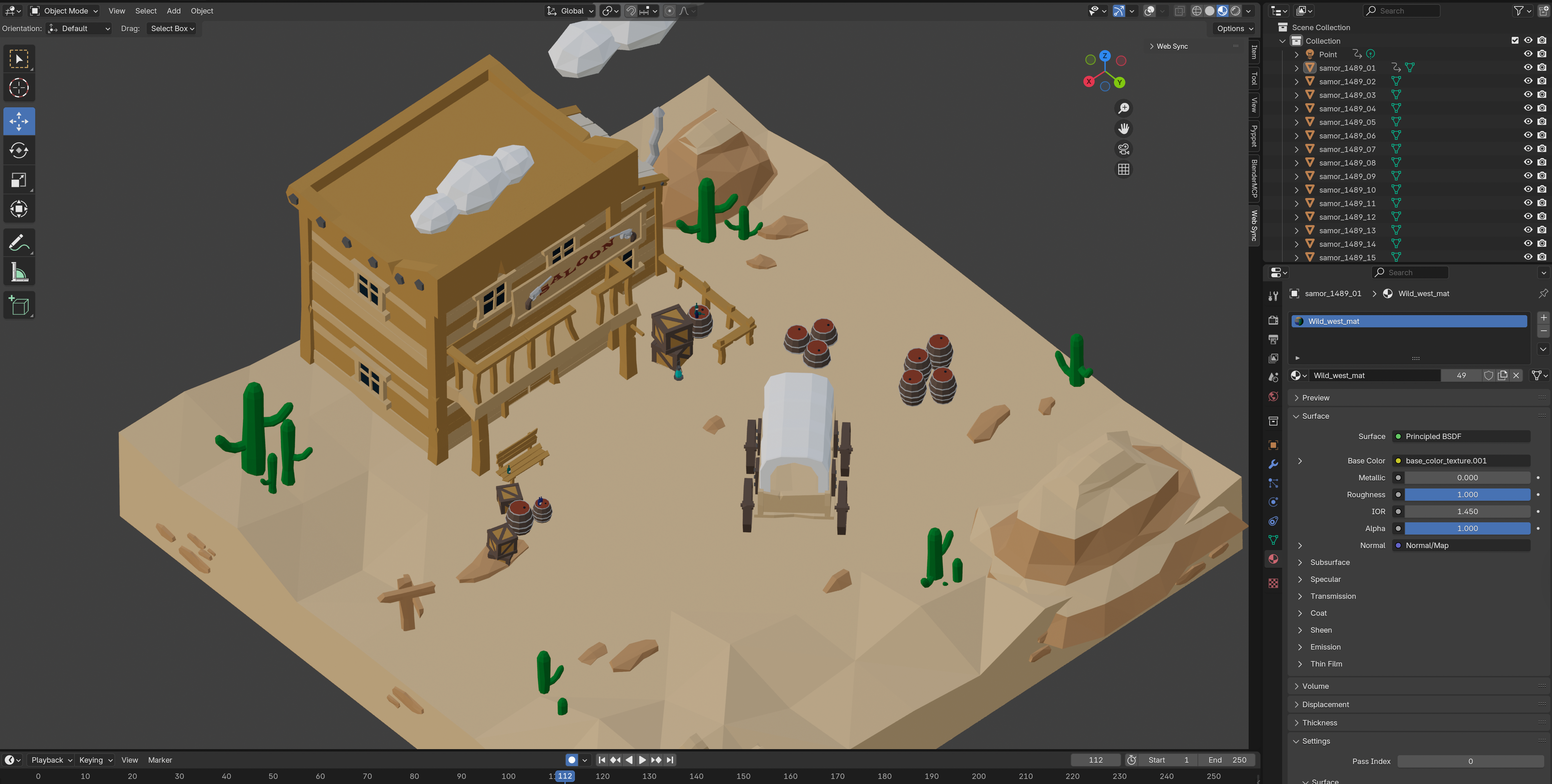}
    \end{minipage}
    
    \caption{
        (Top) Multi-Agent Co-Modeling Interface. 
        (Bottom, left to right) Examples of: manual bi-directional object editing between Blender and Three.js, wireframe mode, and Blender scene being synced in real time.
    }
    \label{fig:multi_agent_interface}
\end{figure}

\section{Experiments and Evaluation}

\subsection{Tasks and Setup}

We curated a set of modeling tasks varying in both scale and type, designed to test the system’s adaptability and generalization. To standardize evaluation criteria and reduce the complexity of modeling operations, we adopted a low-poly style across all asset collection and model generation processes. Experiments were conducted on a MacBook Pro equipped with an Apple M4 Pro processor and 48\,GB of memory.

\begin{table}[hbt!]
\centering
\small
\setlength{\tabcolsep}{3pt}
\renewcommand{\arraystretch}{1.12}
\begin{tabularx}{\textwidth}{p{1.7cm} p{0.2cm} *{3}{>{\centering\arraybackslash}X}}
\toprule
 &  & \textbf{Task 1} & \textbf{Task 2} & \textbf{Task 3} \\
\midrule
\textbf{Category} & & Primitive & Primitive & Primitive \\
\textbf{Scale} & & Small & Medium & Large \\
\textbf{Task} & & Cake & Table & Car \\
\textbf{Initial Prompt} & &
Model a low-poly birthday cake with visible layers and candles. &
Create a low-poly square dining table with four legs. &
Construct a low-poly classic car using only primitive shapes. \\
\multirow{3}{*}{\textbf{Iterations}} & \textbf{1} &
\includegraphics[width=\linewidth,height=0.5625\linewidth,keepaspectratio]{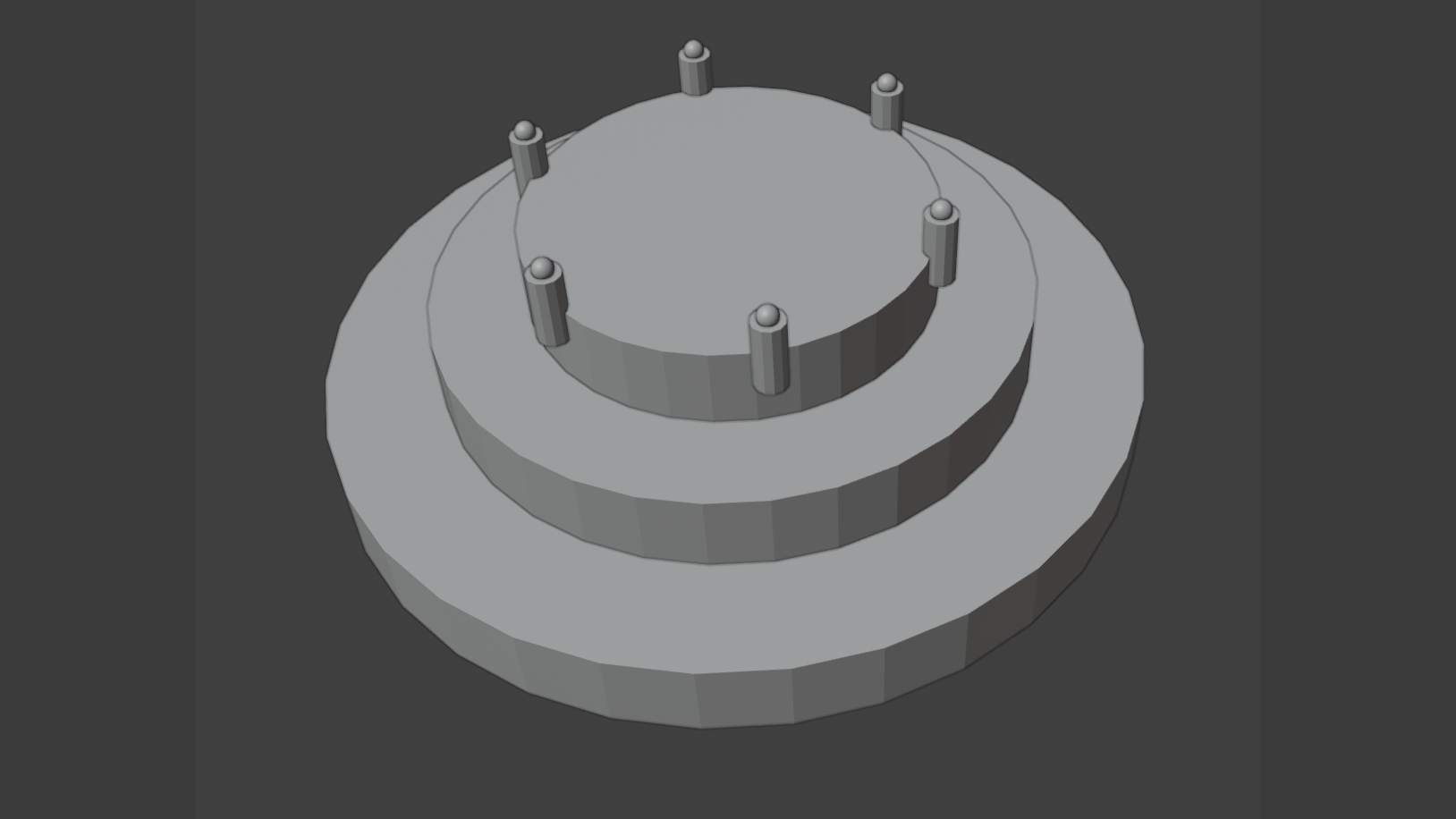} &
\includegraphics[width=\linewidth,height=0.5625\linewidth,keepaspectratio]{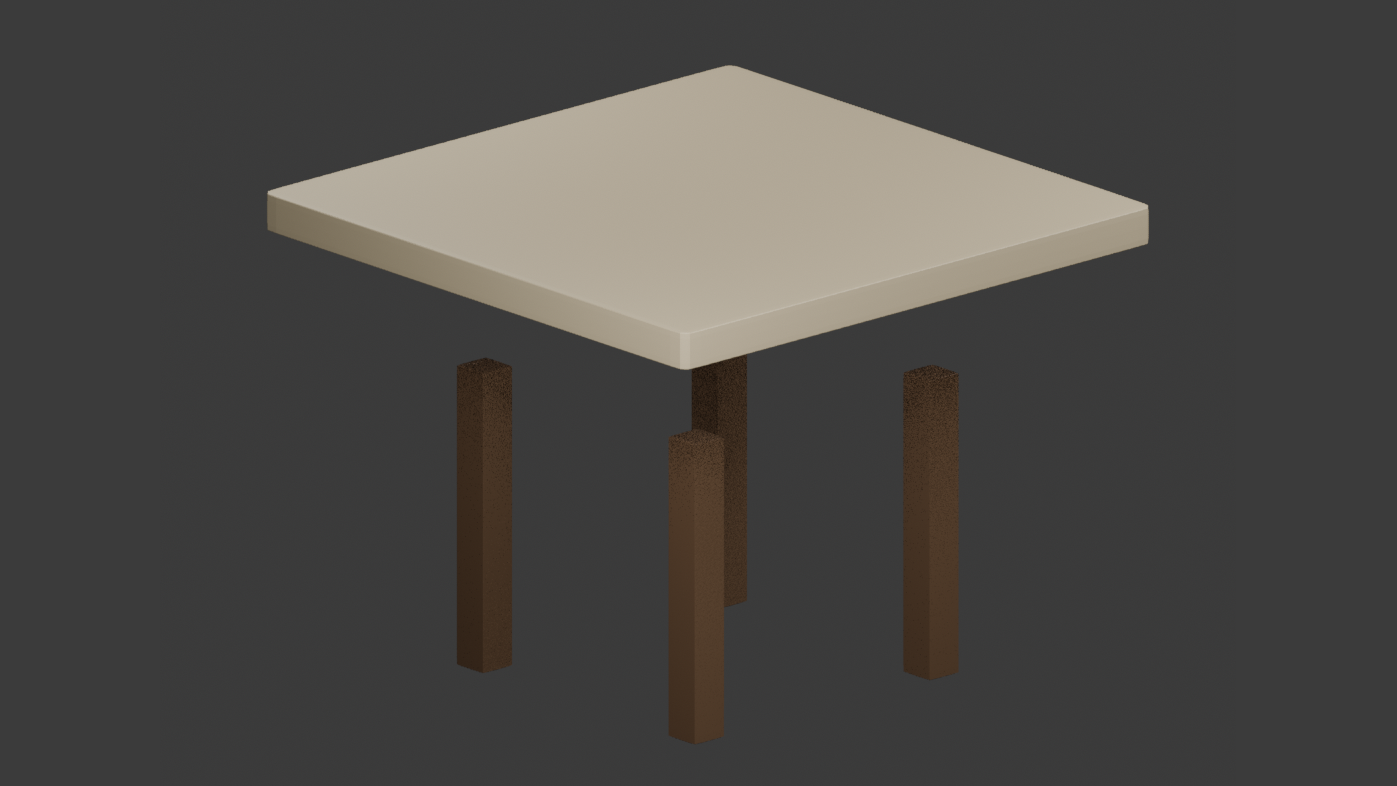} &
\includegraphics[width=\linewidth,height=0.5625\linewidth,keepaspectratio]{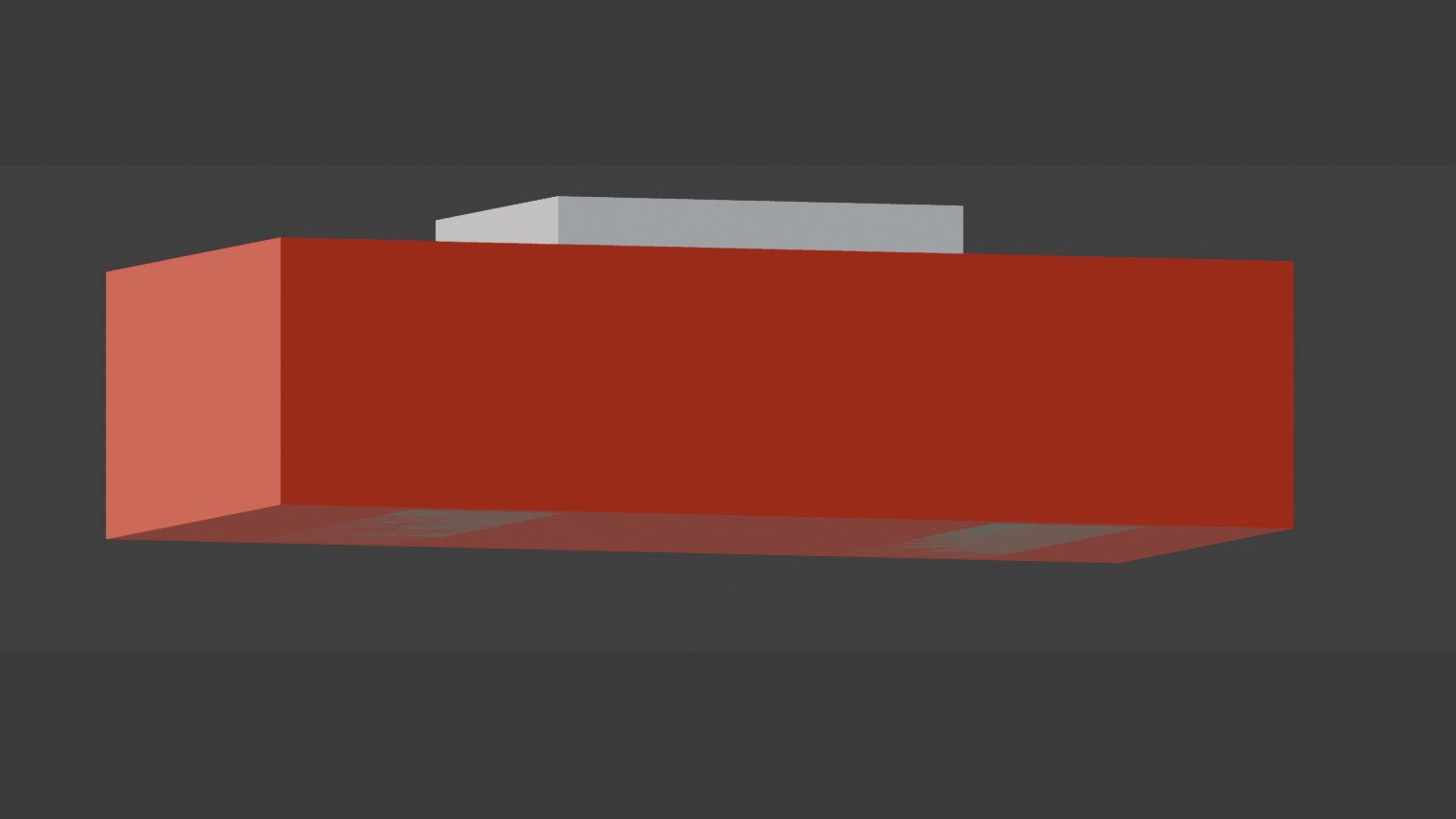} \\
 & \textbf{3} &
\includegraphics[width=\linewidth,height=0.5625\linewidth,keepaspectratio]{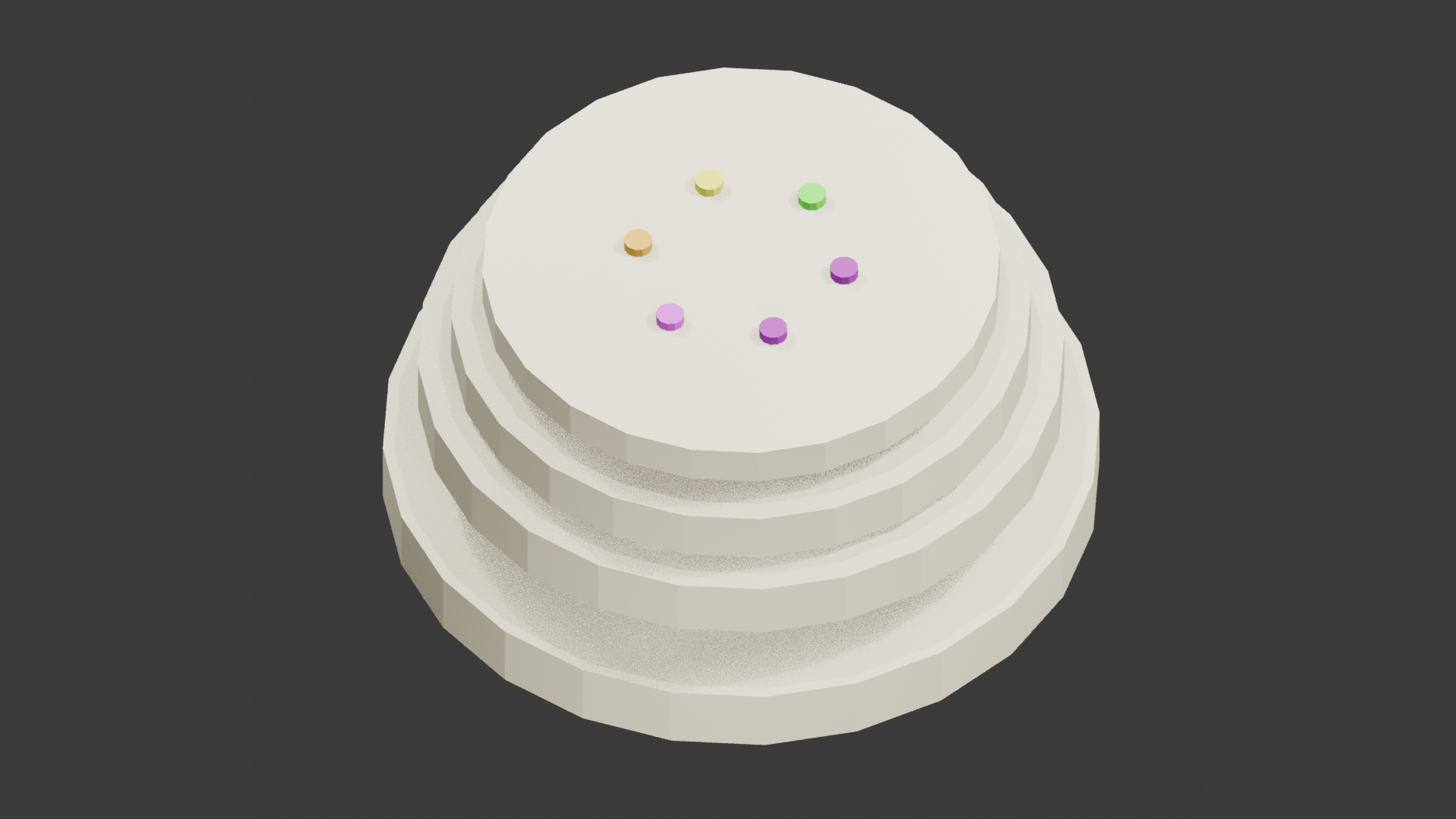} &
\includegraphics[width=\linewidth,height=0.5625\linewidth,keepaspectratio]{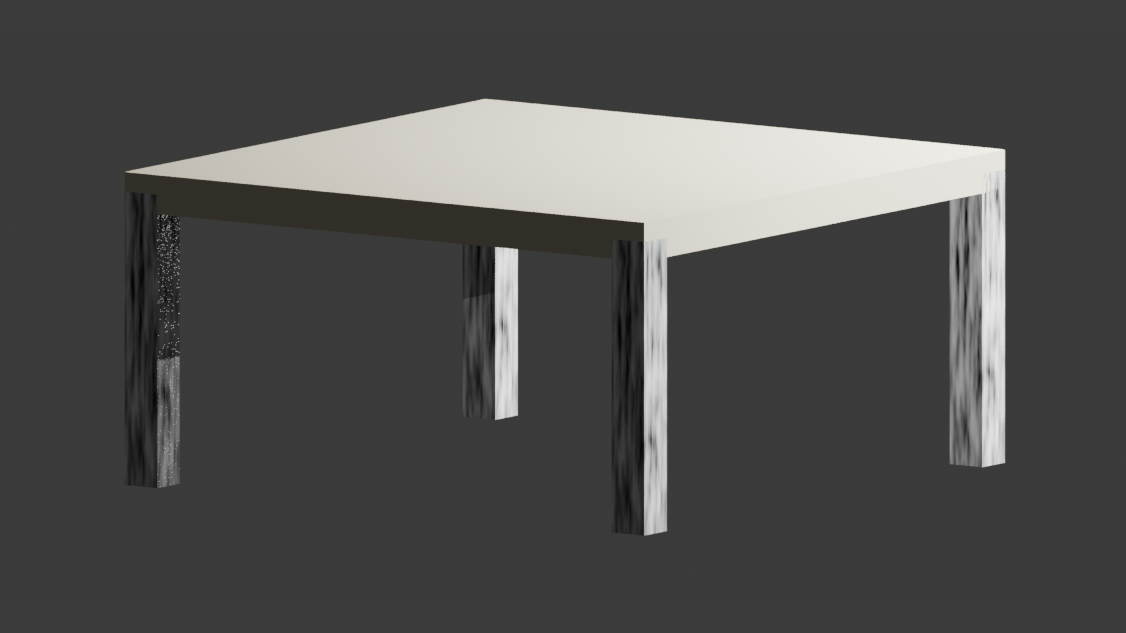} &
\includegraphics[width=\linewidth,height=0.5625\linewidth,keepaspectratio]{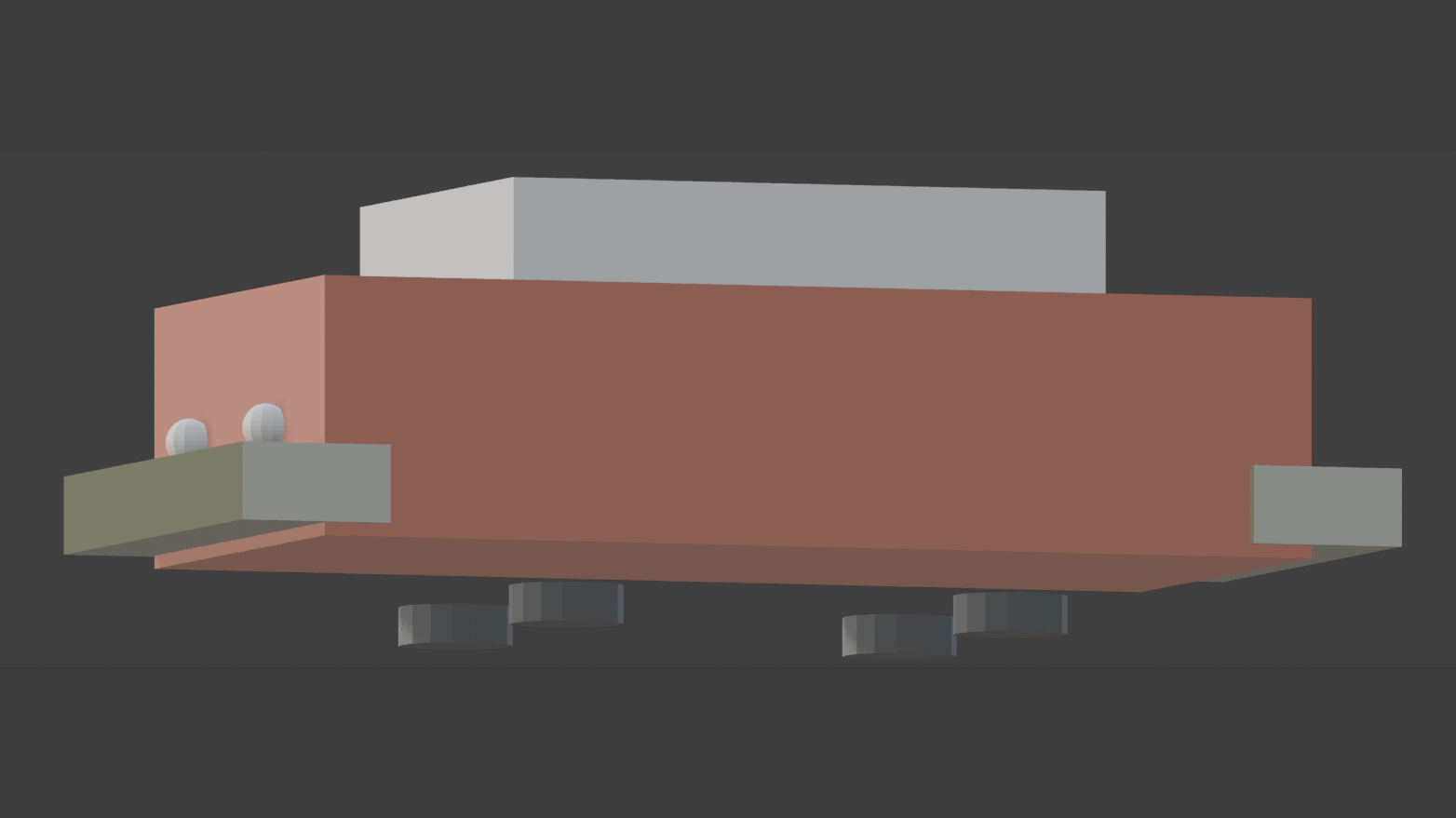} \\
 & \textbf{5} &
\includegraphics[width=\linewidth,height=0.5625\linewidth,keepaspectratio]{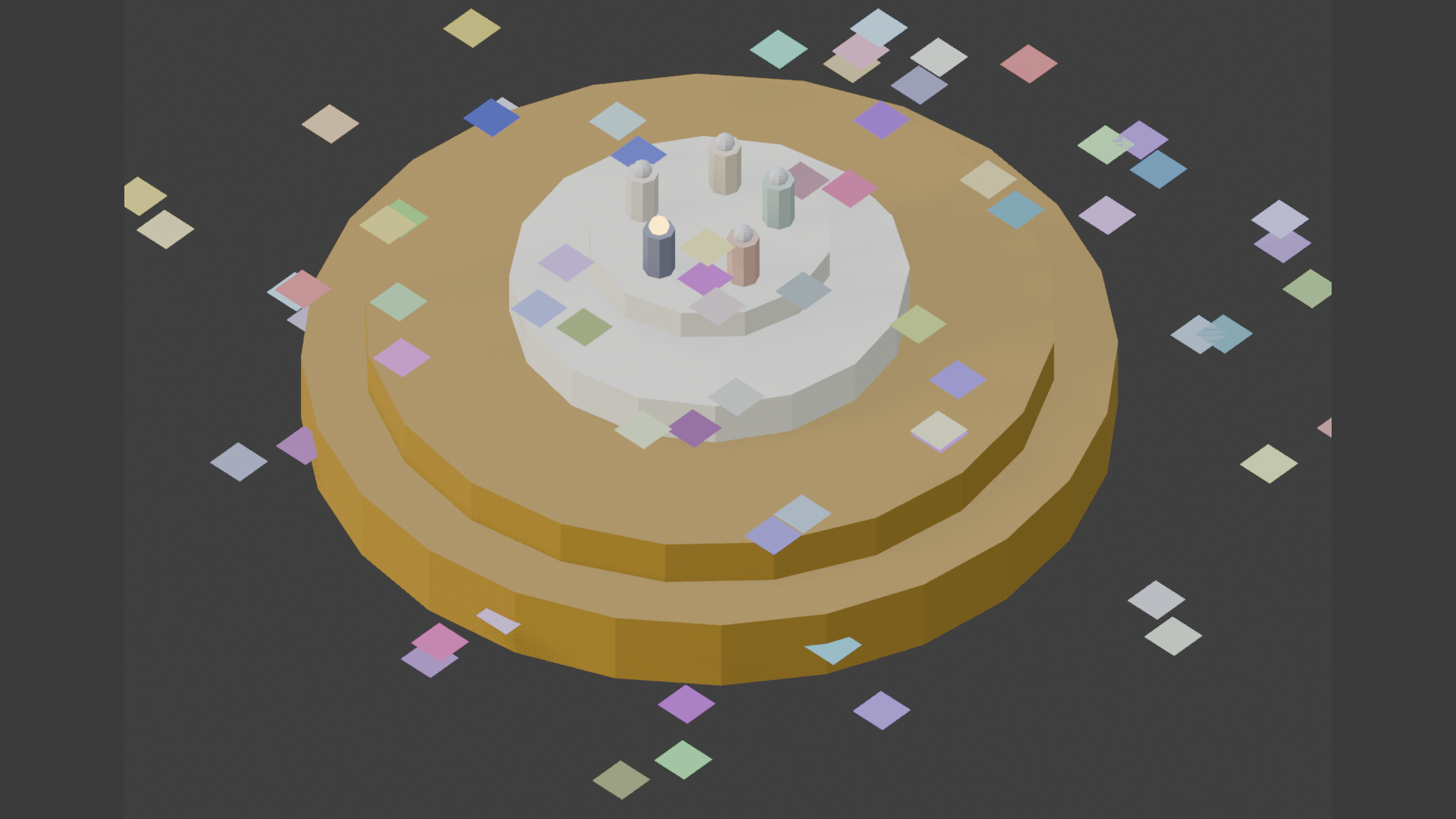} &
\includegraphics[width=\linewidth,height=0.5625\linewidth,keepaspectratio]{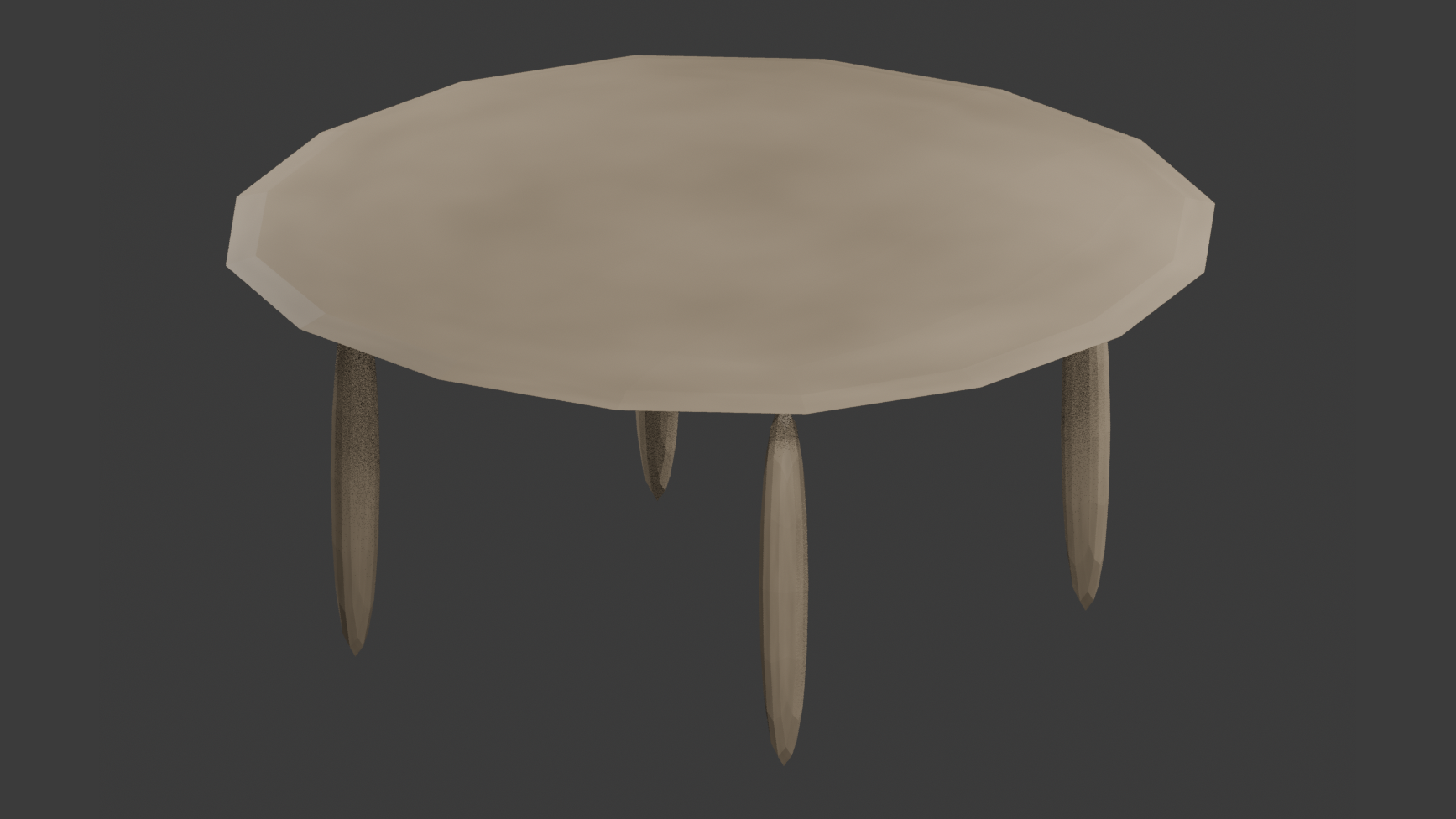} &
\includegraphics[width=\linewidth,height=0.5625\linewidth,keepaspectratio]{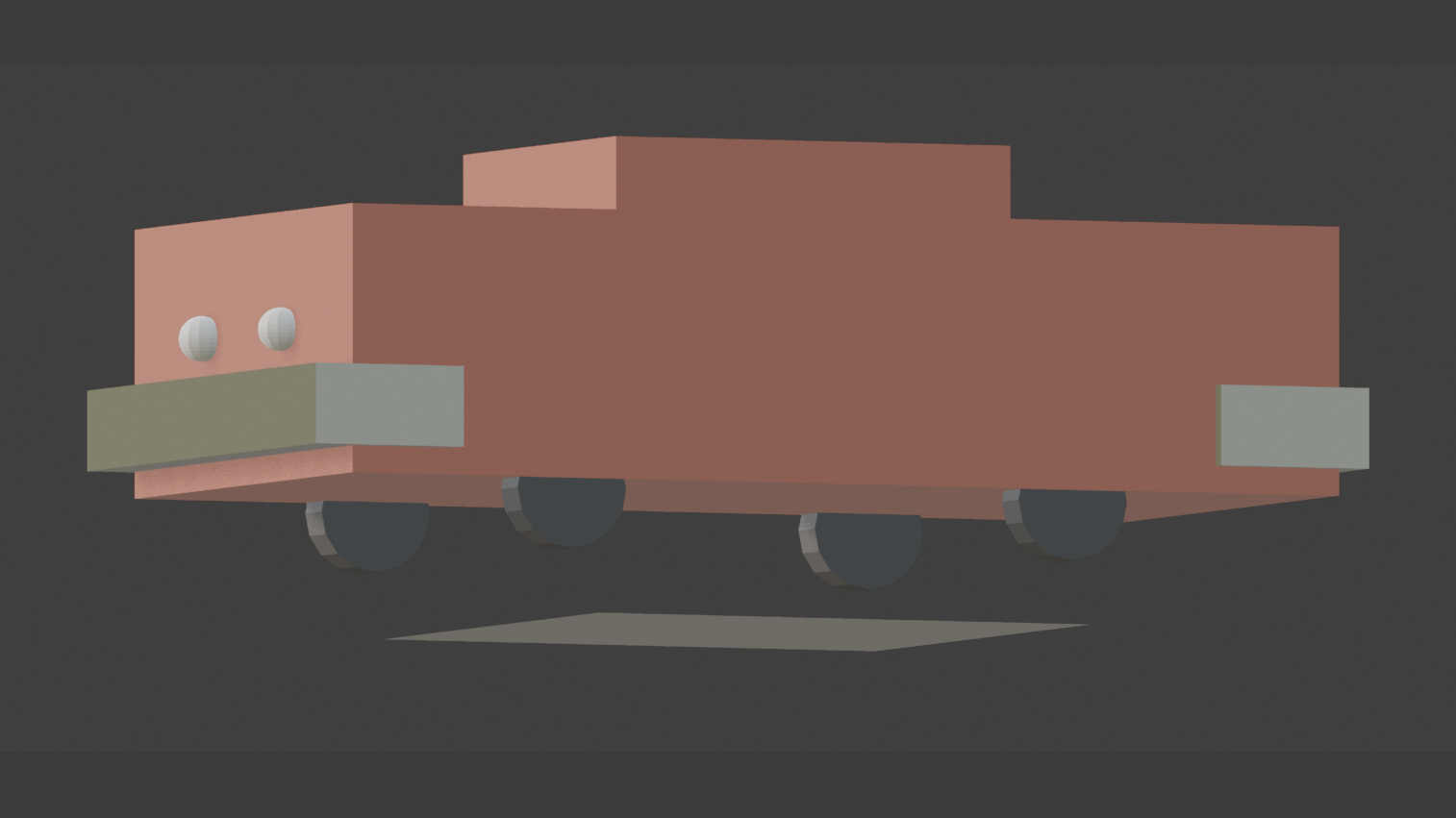} \\
\bottomrule
\end{tabularx}
\caption{Selected Modeling Results on Different Tasks}
\label{tab:task-vertical-matrix}
\end{table}

\subsection{Evaluation Metrics}

We employ a combination of automated and human-in-the-loop evaluation methods across three primary dimensions: \textbf{Geometric Quality}, \textbf{Visual Quality}, and \textbf{Human-in-the-Loop Control}.  
Given the absence of ground truth 3D models for open-ended design tasks, the evaluation follows a hybrid qualitative–quantitative strategy.

\paragraph{Geometric Quality}
Geometric quality is assessed using the following indicators:
\begin{itemize}[noitemsep, topsep=0pt, parsep=0pt, partopsep=0pt]
    \item \textbf{Geometry Count} — The number of distinct geometry objects in the scene recorded after each modeling task.
    \item \textbf{Vertex Count} — The total number of vertices in the scene at each stage. This metric estimates mesh complexity and ensures adherence to low-poly standards, thereby avoiding both overfitting (excessive detail) and underfitting (over-simplification).
    \item \textbf{Similarity} — The similarity between consecutive modeling stages, measuring the extent of scene changes across iterations.
\end{itemize}

\paragraph{Visual Quality}
Visual quality is evaluated through a 5-point Likert scale by three independent human raters using two criteria:
\begin{itemize}[noitemsep, topsep=0pt, parsep=0pt, partopsep=0pt]
    \item \textbf{Task Alignment Score} — Measures the extent to which the generated model fulfills the specified task objective, considering completeness and functional accuracy.
    \item \textbf{Aesthetic Score} — Assesses the visual appeal of the output, considering proportional balance, stylistic consistency, simplicity, and realism.
\end{itemize}
In the actual small-scale experiments, task alignment was positively correlated with the aesthetic quality of the results. Therefore, at this stage, we averaged the two human raters' scores to obtain an overall visual quality score.

\begin{figure}[hbt!]
    \centering
    \includegraphics[width=1\linewidth]{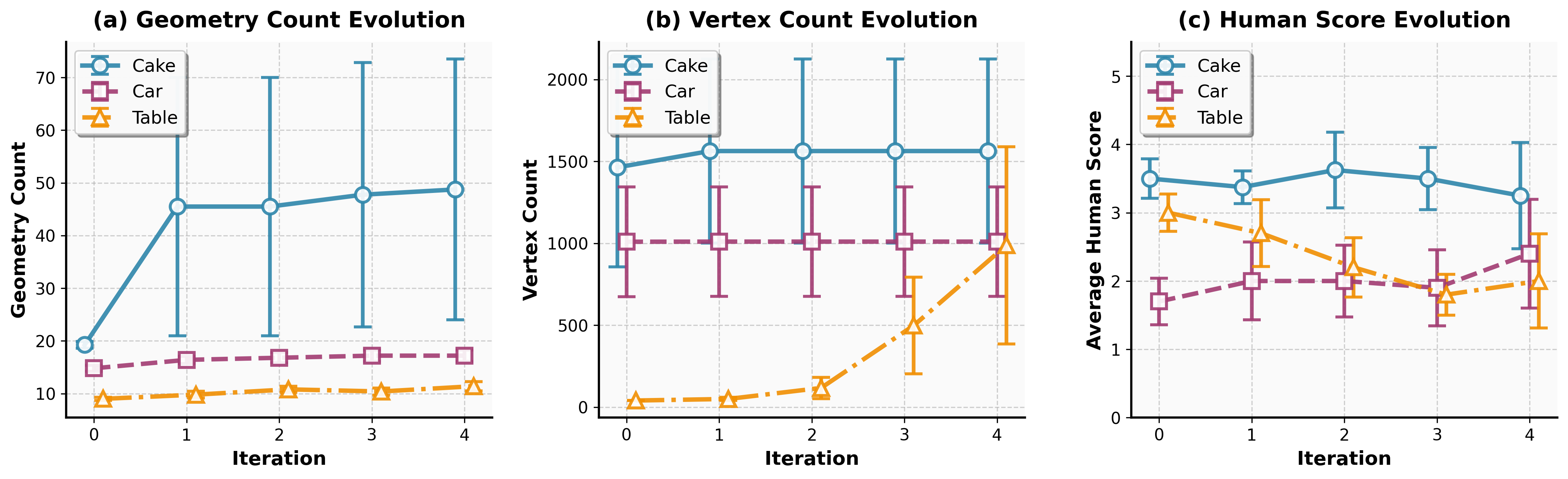}
    \caption{Geometric Quality Analysis in Autonomous Modeling Tasks}
    \label{fig:placeholder}
\end{figure}


\newpage
\section{Discussion \& Future Works}

\subsection{Efficacy of the Self-Reflective Framework}
Across many small to medium-scale modeling scenarios, our self-reflective planner–actor–critic framework achieved higher modeling quality than a single-agent baseline, as shown in the first iteration. Improvements were most evident in geometric proportions, fine-grained adjustments, and material assignment. We observe the planner agent decomposes multi-step modeling goals into executable subtasks that can be completed with short code snippets, which reduces hallucinations and error rates. Therefore, the actor execute more reliable in more complex scenes. The critic inspects scene metadata and screenshots to generate a target-relevance score, influencing the planner’s next decision. Our finding indicate the self-reflective planner–actor–critic pipeline is a viable avenue for automated modeling.

\subsection{Emergent Behaviors and Human Oversight}
Human participation increases controllability and helps the system escape local optima. In practice, users can introduce creative directives during execution, such as switching a square table to a round table, or change materials to a wood color. This human-in-the-loop reduce failure and inject essential human creativity in the agent-augmented modeling.

\subsection{Limitations and Future Improvements}
There are several limitation of our current system. First, due to the limitation of total amount of tools allowed in MCP usage, Python scripts driving Blender can fail due to syntax or logic errors. Second, the actor does not always incorporate the critic’s feedback, and after multiple iterations (often three), the modeling quality degrades or converge, making further improvements difficult. Third, the code execution approach often rely on modeling with primitive, which limit expressiveness. Finally, while human intervention improves outcomes, its impact can be constrained or ignored by agents.

In future work, we will refine the MCP tool reliability by boarden types of tools, and introduce pre-execution validation and automatic repair/retry in the workflow. We will fine-tune prompt and enhanced human-in-the-loop control to better guide the agent convert idea to models.

\newpage
\bibliographystyle{plainnat}
\bibliography{references}  

\newpage


\appendix

\section{Technical Appendices and Supplementary Material}
Technical appendices with additional results, figures, graphs and proofs may be submitted with the paper submission before the full submission deadline (see above), or as a separate PDF in the ZIP file below before the supplementary material deadline. There is no page limit for the technical appendices.

The source code used in the experiments is available at: \url{https://drive.google.com/drive/folders/12Pf5axYPWa5WYOnwu3p0ie6M4BxhHnmj?usp=drive_link}

The prompt definition and the screenshots for agent creation experiment are included in the following pages.

\newpage

\subsection{Planning Agent Prompting}

The Planning Agent is responsible for creating detailed execution plans based on user requests. Its prompting system is found in both \texttt{blender\_agents.py} and \texttt{agent.py}.

\subsubsection{Core System Prompt Template}

\begin{lstlisting}[language=Python, basicstyle=\tiny\rmfamily]
system_prompt = f"""You are a Blender planning agent. Create step-by-step plans for 3D modeling.

AVAILABLE BLENDER MCP TOOLS (USE ONLY THESE):
{tools_description}

{revision_context}

IMPORTANT RULES:
1) Use ONLY the tools listed above. Do NOT invent tool names.
2) Respect required parameters shown for each tool.
3) If you need to run Blender Python, use `execute_code` or `execute_blender_code`
and include a COMPLETE script in parameters.code (no prose, no markdown fences, no print-only placeholders).
- Start at column 0, import bpy, and perform real ops (bpy.ops.* / bpy.data.*).
- Give created objects explicit names so later steps (e.g., get_object_info) can reference them.
4) Always end with a `get_viewport_screenshot` step to verify results.

You MUST call the create_plan function with a detailed plan.
"""
\end{lstlisting}

\subsubsection{Dynamic Context Injection}

The Planning Agent dynamically incorporates several contextual elements:

\begin{enumerate}
    \item \textbf{Tool Discovery}: Real-time discovery of available MCP tools
    \item \textbf{Revision Context}: Integration of critique feedback for iterative improvement
    \item \textbf{Scene State}: Current Blender scene objects for incremental modeling
    \item \textbf{User Feedback}: Direct incorporation of human feedback from review phases
\end{enumerate}

\subsubsection{Iterative Planning Prompts}

For subsequent iterations, additional context is provided:

\begin{lstlisting}[language=Python, basicstyle=\tiny\rmfamily]
if state.get("iteration_count", 0) > 0:
    system_prompt += """
WHEN ITERATING:
- Assume the Blender scene already contains the previous attempt.
- Do NOT recreate objects that already exist; modify them instead.
- First call get_scene_info / get_object_info to discover existing objects.
- Use deterministic names and reuse them across iterations.
- Only create missing parts that the critique called out.
- End with get_viewport_screenshot.
"""

system_prompt += """
NAMING:
- Derive object names from semantic roles (e.g., 'Main_Body', 'Left_Wheel', 'Window_Panel').
- Reuse these names across iterations; only create missing parts.
- Never clear or reset the scene.
"""
\end{lstlisting}

\subsection{Blender/Actor Agent Prompting}

The Blender Agent executes the generated plans using MCP (Model Context Protocol) tools. Its prompting focuses on reliable tool execution and error handling.

\subsubsection{Execution System Prompt}

\begin{lstlisting}[language=Python, basicstyle=\tiny\rmfamily]
system_prompt = """You are a Blender automation agent executing a specific plan.
You have access to Blender MCP tools. Follow the plan exactly.

For each step:
1. Use the specified tool with the given parameters
2. Verify the action completed successfully
3. Report the result

Available tools: {tools}

Current plan step: {current_step}
"""
\end{lstlisting}

\subsubsection{Dynamic Code Synthesis}

The system includes an advanced code synthesis mechanism for Blender operations:

\begin{lstlisting}[language=Python, basicstyle=\tiny\rmfamily]
async def synthesize_blender_code(action: str) -> str:
    llm = ChatOpenAI(model="gpt-4.1", temperature=0)
    sys = (
        "Return ONLY executable Python for Blender (no markdown fences/comments/prints). "
        "Start at column 0. import bpy. Use bpy/bpy.ops. "
        "NEVER clear or reset the scene (no select_all/delete, orphan_purge, open_mainfile, read_homefile, new scene). "
        "NEVER delete objects; if something must be hidden or removed from view, move it to x+=1000 instead. "
        "Be IDEMPOTENT: if an object with the intended name exists, modify it; otherwise create it. "
        "Use explicit names; reuse them. "
        "When adjusting size, use obj.dimensions or obj.scale consistently and keep transforms sane. "
        "If the requested action is a modification but the target object is not found, do nothing."
    )
    
    ask = f"Write Blender Python to accomplish this action: {action}"
    resp = await llm.ainvoke([SystemMessage(content=sys), HumanMessage(content=ask)])
    return process_generated_code(resp.content)
\end{lstlisting}

\subsection{Critic Agent Prompting}

The Critic Agent analyzes execution results and screenshots to provide feedback for iterative improvement.

\subsubsection{Analysis System Prompt}

\begin{lstlisting}[language=Python, basicstyle=\tiny\rmfamily]
system_prompt = """You are a 3D modeling critique expert. Analyze the Blender viewport screenshot and execution results.
Consider the original plan's success criteria and provide constructive feedback.

You MUST call the analyze_result function with your analysis."""

critique_context = f"""
Original Request: {state.get('user_request', '')}

Success Criteria:
{chr(10).join(f'- {c}' for c in plan.get('success_criteria', []))}

Execution Summary:
{state.get('execution_summary', '')}

Please analyze the screenshot and provide feedback on:
1. Whether the success criteria were met
2. Quality of the 3D model/scene
3. Any issues or improvements needed
4. Whether another iteration is needed
"""
\end{lstlisting}

\includepdf[pages=-]{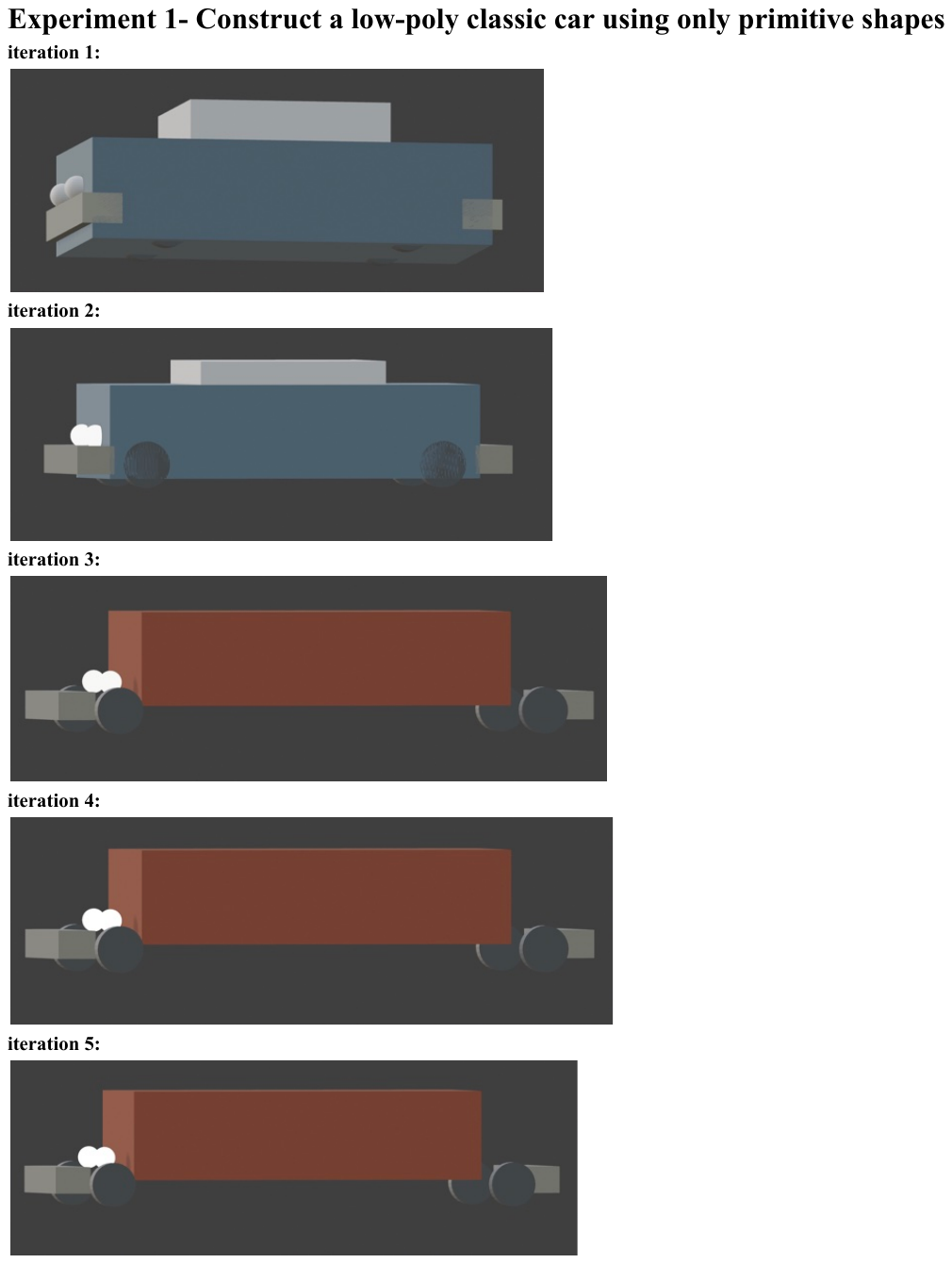}

\end{document}